

\magnification=\magstep1
\font\titlefont=cmr10 scaled\magstep3  
\font\secfont=cmbx12 
\newbox\leftpage \newdimen\fullhsize \newdimen\hstitle \newdimen\hsbody
\hoffset=0.0truein \voffset=0.20truein \hsbody=\hsize \hstitle=\hsize
\tolerance=1000\hfuzz=2pt \baselineskip=20pt plus 4pt minus 2pt
\global\newcount\meqno \global\meqno=1
\def\eqn#1#2{\xdef #1{(\the\meqno)}\global\advance\meqno by1 $$#2\eqno#1$$}
\global\newcount\refno \global\refno=1 \newwrite\rfile
\def\ref#1#2{[\the\refno]\nref#1{#2}}%
\def\nref#1#2{\xdef#1{[\the\refno]}%
\ifnum\refno=1\immediate\openout\rfile=refs.tmp\fi%
\immediate\write\rfile{\noexpand\item{[\the\refno]\ }#2}%
\global\advance\refno by1}

\global\newcount\figno \global\figno=1 \newwrite\ffile
\def\fig#1#2{(\the\figno)\nfig#1{#2}}%
\def\nfig#1#2{\xdef#1{(\the\figno)}%
\ifnum\figno=1\immediate\openout\ffile=figs.tmp\fi%
\immediate\write\ffile{\noexpand\item{\sl Fig.(\the\figno)\ }#2}%
\global\advance\figno by1}

\nopagenumbers \hsize=\hsbody \pageno=0 ~ \vfill
\centerline{\titlefont Conductance Distributions in Random Resistor Networks;}
\centerline{\titlefont Self Averaging and Disorder Lengths}
\medskip
\vfill\centerline{{\sl Rafael F. Angulo and Ernesto Medina}}
\medskip
\centerline{Coordinaci\'on de Investigaci\'on B\'asica}
\centerline{Intevep S.A.,  Apartado 76343}
\centerline{Caracas 1070A, Venezuela}
\vfill\centerline{\bf ABSTRACT}\nobreak\medskip\nobreak\par
{The self averaging properties of the conductance $g$  are explored in Random
Resistor Networks (RRN) with a broad distribution of bond strengths $P(g)\sim
g^{\mu -1}$. The
RRN problem is cast in terms of simple combinations of random variables on
hierarchical lattices. Distributions of
equivalent conductances are estimated numerically on hierarchical lattices as a
function
of size $L$ and the distribution tail strength parameter $\mu$. For
networks above the percolation threshold, convergence to a Gaussian basin is
always the case, except in
the limit $\mu\rightarrow 0$. A {\it disorder length} $\xi_D$ is
identified, beyond which the system is effectively homogeneous. This length
scale
diverges as $\xi_D\sim |\mu |^{-\nu}$, ($\nu$ is the regular percolation
correlation length exponent) when the
microscopic distribution of conductors is exponentially wide ($\mu\rightarrow
0$). This implies that exactly
the same critical behavior can be induced by geometrical disorder and by strong
bond disorder with
the bond occupation probability $p \leftrightarrow \mu$. We find that only
lattices
at the percolation threshold have renormalized probability distributions in a
{\it Levy-like} basin.
At the percolation threshold the disorder length diverges at a critical tail
strength $\mu_c$ as
$|\mu-\mu_c|^{-z}$ with $z\sim 3.2\pm 0.1$, a new exponent. {\it Critical path
analysis} is used in a
generalized form to give the macroscopic conductance in the case of lattices
above $p_c$.}
\vfill\eject\footline={\hss\tenrm\folio\hss}
\vskip .3in \pageno=1

Scaling properties of transport quantities in random media have attracted
great interest for some time. Electric conduction on random networks, in
particular, is of great relevance
because it may be regarded as a simple qualitative archetype of other, more
complex and less
well understood problems in disordered systems. Such physical systems include
the classical
problem of flow in porous media\ref\rFPM{C. Baudet, E. Charlaix, E. Clement, E.
Gyron,
F. Hulin and C. Leroy in {\it Scaling Phenomena in Disordered Systems}, R. Pynn
and A. Skeltorp editors
,NATO ASI Series (1985). }, diffusion in a random environment\ref\rBouch{J. P.
Bouchaud and A. Georges,
Phys. Rep. {\bf 195}, 127 (1990).} and even the
quantum problem of hopping conduction in dirty semiconductors\ref\rSE{B. I.
Shklovskii
and A. L. Efros, {\it Electronic Properties of Doped Semiconductors},
Springer-Verlag, Berlin
(1984).}\ or metal insulator composites\ref\rMIComp{See {\it Physical Phenomena
in Granular Materials},
Materials Research Society Symposium Proceedings, Vol 195, edited by G. D.
Cody, T. H. Geballe and P. Sheng (MRS
Pittsburgh, Pensylvania, 1990).}, among others. These systems
can be modeled on a random network where individual bonds
represent a `resistor' whose properties depend on the quantity which flows
through it. In the case
of flow in porous media this resistor represents a microscopic permeability or
{\it permeance} (in
analogy with {\it conductance} to indicate that it may not be an intensive
quantity). For hopping
conduction, on the other hand, the resistor
involves the computation of the overlap integral between impurities, and
quantum effects
are taken into account\rSE. The physics involved in the computation of the
resistors
determines an associated probability distribution function (PDF) for values on
the equivalent resistor
network. The resulting distribution for hopping conduction is exponentially
wide ($r=r_0 exp(-\epsilon)$
with $\epsilon$, say, uniformly distributed), while in the case of flow through
porous media in the capillary regime
it has been argued that the distribution
has a power law component and a large permeability cutoff\ref\r{P. G. Toledo,
H. T. Davis, and L. E. Scriven
Physica A {\bf 185}, 228 (1992).}. In the same
fashion, continuum percolation results in a random network with a power law
distribution of
resistance $P(r)\sim 1/r^{\mu +1}$\ref\rHFS{B. I. Halperin, S. Feng, and P. N.
Sen,
Phys. Rev. Lett. {\bf 54}, 2391 (1985).}. In addition to the {\it statistical}
disorder discussed
above, one can also introduce {\it geometrical} disorder by allowing voids in
the network with a certain
probability. Networks
with a broad distribution of bond values at the percolation threshold have been
extensively studied numerically
and by field-theoretical techniques\ref\rSelI{J. P. Straley,
J. Phys. C {\bf 15}, 2333 (1982); {\bf 15}, 2343 (1982) 2343; B. I. Halperin,
S. Feng, and P. N. Sen, Phys. Rev.
B {\bf 35}, 197 (1987).}\ref\rLT{T. C. Lubensky
and A. M. S Tremblay, Phys. Rev. B {\bf 34}, 3408 (1986); see also T. Schlosser
and H. Spohn, J. Stat. Phys.
 {\bf 69}, 955 (1992).}\ref\rMacII{J. Machta,
Phys. Rev. B {\bf 37}, 7892 (1988).}.

Once the distribution of the elementary resistors is set, as discussed above,
one is interested
in computing the macroscopic properties of such networks. How does the
probability distribution
of the equivalent conductance `renormalize' as the network is rescaled? What is
the effective
support for transport, given the elementary distribution of resistors? In brief
how do the random variables {\it combine} and thus `interact' with the
microscopic probability
distribution (MPD)
to give a macroscopic result. In this work we explore the scaling properties of
conduction
on {\it hierarchical networks} where the problem can be simply and explicitly
stated in terms of a non-linear
combination of random variables. This approach is very attractive since it
dwells on the problem
of stable limit distributions for combinations of random variables (other than
additive) for which
there are few rigorous results. This approach has been adopted recently by
Derrida and Griffiths\ref\rSelII{B. Derrida
and R. B. Griffiths, Europhys. Lett. {\bf 8},
111 (1989)}, Halpin-Healy\ref\rHH{T. Halpin-Healy, Phys Rev. A {\bf 42}, 711
(1990)}\ and
Roux et al\ref\rRHSLP{S. Roux,
A. Hansen, L. R. da Silva, L. S. Lucena and R. B. Pandey, J. Stat. Phys.
{\bf 65}, 183 (1991).}\ in the context of directed polymers in a random medium
and by Bouchaud, Le
Doussal and Georges in a series of papers reviewed in ref.\rBouch\ in the
context of anomalous diffusion.

The paper is organized as follows: In section I we
discuss hierarchical lattices and how resistor network composition laws are
easily
implemented. We then show the results for the evolution of the PDF as a
function
of system size for hierarchical networks at, and above the percolation
threshold.
The existence of two basins of attraction is obtained as a function of the
MPD tail strength parameter. In section II we discuss the existence of a
disorder
length $\xi_D$ below which strong heterogeneities persist, and derive the
conductance at this scale by   {\it critical path analysis} introduced by
Ambegaokar, Halperin and
Langer (AHL) \rSE\ref\rAHL{V. Ambegaokar,
B. I. Halperin and J. S. Langer, Phys. Rev. B {\bf 4}, 2612 (1971)}. This
length scale determines the
crossover distance over which the system self averages. We also show the
divergence of $\xi_D$ as
one approaches $P(g)\sim 1/g$ and determine the associated critical exponent
numerically and using
simple analytical arguments. We find that this exponent is exactly that of
regular percolation, and
establish a mapping between the probability $p$ of regular percolation and the
tail strength
parameter $\mu$.  In section III we point out the parallelism between Levy
limit theorems and
limit distributions found in random resistor networks. Analogies to other
disordered systems are
also discussed.

\bigskip

{\noindent{\secfont I. Hierarchical Lattices. }}

Hierarchical lattices are self similar structures generated by an iterative
procedure that
produces the lattice at order $m+1$ from order $m$, by substituting every bond
by
a chosen motif (see fig.\fig\fOneA{Recursive construction of a hierarchical
lattice: Each bond at
a given generation $m$ is converted to a chosen motif. (a) The Berker lattice
$d_e=2$, (b) the
ARC lattice of $d_e=\log4/\log3$.}). With this procedure one obtains lattices
of different
effective dimensions $d_e$ and connectivities, which have proven useful
in exposing the qualitative behavior of low dimensional systems\ref\rBO{
A. N. Berker and S. Ostlund, J. Phys. C {\bf 12}, 4961 (1979).}. Fig.\fOneA\
shows  the so called
`Berker lattice' of effective dimension $d_e=2$  and the Arcangelis, Redner and
Coniglio (ARC)
lattice\ref\rARC{L. Arcangelis, S. Redner, and
A. Coniglio, Phys. Rev. B {\bf 31}, 4725 (1985).}\ frequently used as a model
for percolating
backbones ($d_e=\log4/\log3$.)

Bonds on the hierarchical lattices are assigned conductances $g$,
independently chosen from the probability distribution
\eqn\ePDFg{P(g)=|\mu|g^{\mu-1}
{}~{\rm with}~\cases{ 0\leq g \leq 1 &if $\mu>0$\cr g>1 &if $\mu<0$\cr}.}
The ranges assigned to the conductance $g$ secure that the distribution is
normalized for the
given values of the tail strength $\mu$. Various limiting behaviors can be
achieved in
eq.\ePDFg\ by varying $\mu$, namely: a) $~\mu >>1$ corresponds to
$P(g)\rightarrow \delta(g-1)$,
b) $~\mu =1$ to a flattop distribution between $0<g<1$,
c) $~0<\mu<1$ corresponds to continuum percolation, and d)$~\mu <0$ to
algebraic tails for $g \rightarrow
\infty$. In addition one can study the limit e) $~\mu\rightarrow 0^{\pm}$ which
corresponds to the relevant distribution for dirty semiconductors in the
hopping regime\rAHL.

Some background and notation for transport and percolation exponents is in
line:
For infinite systems near $p_c$ above the threshold, the conductivity
$\sigma$ behaves as $\sigma \sim (p-p_c)^t$,
where $t$ is the conductivity exponent. In the percolation problem a
characteristic length
$\xi \sim (p-p_c)^{-\nu}$ of geometrical origin emerges. The correlation length
exponent
$\nu$ is defined by the previous expression.
Below this length scale the system is strongly inhomogeneous and has a fractal
structure, and thus intensive quantities such as the conductivity scale with
the length
$L$ in a nontrivial manner. Beyond this scale the system becomes effectively
homogeneous, and intensive quantities take their macroscopic value ( the
macroscopic average conductance
$\langle G\rangle=\sigma L^{d-2}$ with $\sigma$ independent of $L$.) It is an
accepted
theoretical result\rSelI\rLT\rMacII\ that when the PDF of conductors
corresponds to the continuum percolation distribution and the underlying
lattice is at $p_c$, one finds a regime where
universal {\it lattice} exponents are found, while values of $\mu$ below
a certain threshold $\mu_c$ yield tail strength dependent exponents. This
behavior is summarized by the expression
$t=max\{t_0,t(\mu)\}$,
where $t_0=(d-2)\nu+\phi_0$ and $t(\mu)=(d-2)\nu+1/\mu$.

Hierarchical lattices of length $L$ ranging from $2$ to $16384$ with random
microscopic
conductances obtained from eq.\ePDFg, are
generated for selected values of $\mu$. The equivalent
conductance $G$ of each lattice is then calculated exactly. More than 10000
realizations of randomness are generated to produce a sample from which a
renormalized
distribution of equivalent conductances can be estimated. Our results are the
following:

 1) {\it For lattices above the percolation threshold},
we find a rapid convergence  of the MPD to a sharp Gaussian for all $\mu\ne 0$,
as shown in
fig.\fig\fTwoA{Convergence of a broad microscopic distribution towards a
Gaussian as $L$ is increased. a) Shows
the case of the Berker lattice of $d_e=2$ and $\mu=0.4$ and b) the Berker
lattice for $d_e=3$ and $\mu=0.3$. The
size of the lattices used are: squares L=16, filled circles L=256, triangles
L=16384 . The inset shows
a fit to as Gaussian distribution.}. The conductance
is then a {\it self averaging} quantity in the sense that, as the
network increases in size the conductance approaches a limit value with
diminishing
fluctuations. For the conductance one finds that $\langle G\rangle \sim
G_{typ}$ for
large enough $L$, indicative of a peaked distribution ($G_{typ}$ stands for
most probable
conductance).

Fig.\fig\fThree{The figure shows the conductance versus the scaling
variable $L/\xi_D$ for a lattice of $d_e=2$, above the threshold.
A crossover length $\xi_D$ and a characteristic conductance $G_{\xi_D}$ are
identified by collapse of
different $\mu$ values. The length $\xi_D$ diverges
when $\mu\rightarrow 0$ as $A_{\pm}|\mu |^{-1.6\pm 0.1}$, where $A_+/A_-=0.6$.
}\ shows the behavior of the
conductance as a function of the length  $L$. Two well defined
regions are apparent; for small networks the conductance drops sharply,
while beyond a given {\it disorder} length $\xi_D$, classical ($G\sim L^{d-2}$)
behavior is recovered. The curves
for different values of $\mu$ can be collapsed by scaling $L$ by $\xi_D$ and
the conductance $G$
by $G_{\xi_D}$ (the conductance at the disorder length). The result is two
universal curves for $\mu
>0$ and $\mu<0$. In the limit $\mu \rightarrow 0$, the collapse indicates that
$\xi_D$ diverges as a
power of the tail strength parameter $\mu$ i.e. $\xi_D = A_{\pm}|\mu|^{-z}$,
as depicted in fig.\fThree. The ratio of the amplitudes is $A_+/A_-=0.6$ and
the exponent $z=1.6\pm 0.1$. The
subscripts on the prefactor indicate the approach from either $\mu >0$ or $\mu
<0$. For
lengths larger than $\xi_D$, the conductivity reaches its macroscopic
value (given $\xi<\xi_D$). One can then write (in $d$ dimensions)
\eqn\eGvsL{\langle G\rangle =\langle G_{\xi_D}\rangle
(L/\xi_D)^{d-2}=\left\langle {G_{\xi_D}\over
\xi_D^{d-2}}\right\rangle L^{d-2}=\sigma L^{d-2},}
which defines the macroscopic {\it conductivity} $\sigma=\langle
G_{\xi_D}/\xi_D^{d-2}\rangle $.
One can go further and derive the form of $\langle G_{\xi_D}\rangle$ using
critical path analysis, valid for $\xi_D$ sufficiently large\rAHL\ref\rTH{S.
${\rm Ty\check{c}}$ and
B. I. Halperin, Phys. Rev. B {\bf 39}, 877 (1989); See also B. I. Halperin,
Physica D
{\bf 38}, 179 (1989) .}. This is done in the next section. For $d_e=3$ collapse
is achieved by scaling
the vertical axis by
$G_{\xi_D}=(1-p_c)^{\nu +1/\mu}\mu^{\nu}$, scaling variable
which will be discussed in the next section. In figure\fig\fFive{Lattice of
bonds carrying
$99\%$ of the current as a
function of $\mu$. The hierarchical lattice generator is conveniently drawn as
shown in the figure. Only the
lengths along the vertical axis in the figure are meaningful}\ we show the
effective support for conduction on
a hierarchical lattice of $d_e=2$ conveniently drawn for the purpose of
illustration (top of figure).
We depict only the bonds supporting $99\%$ of the current on the hierarchical
network as a function of $\mu$. The
disorder length $\xi_D$ derived above is related to the vertical length
of the voids. As disorder increases ($\mu\rightarrow 0$) the current is
essentially carried by a percolation backbone.

In summary no matter how broad the MPD (except  for $\mu=0$), the limiting
behavior
for $P(g)$ on hierarchical networks of $d_e=2,3$ converges to a sharp gaussian.
Geometrical
disorder produced when $p_c<p <1$ seems not to affect the limit behavior of
$P(g)$.  A disorder length $\xi_D$ is found for the convergence to the
`Gaussian' basin, which
defines the length beyond which fluctuations in the conductance decrease around
an average value
$\langle G\rangle=G_{typ}$.

2) {\it For networks at the threshold}, two regimes occur as found by Machta,
Guyer and Moore
(MGM)\ref\rMacIII{J. Machta R. A. Guyer and S. M. Moore, Phys. Rev. B {\bf 33},
4818 (1988).}. For
$\mu>\mu_c\sim0.75$
the system is self-averaging so that a peaked distribution is obtained for
large $L$. In this regime universal lattice
exponents are recovered. We note that the limit distribution for resistors,
although highly peaked, preserves a degree
of skewness (as measured by the third cumulant) as far as we can determine.
This means that the limit PDF
may not approach a Gaussian behavior as found for lattices above the threshold.
On the other
hand for $\mu<\mu_c$  the MPD tails are preserved
(see fig.\fig\fSix{Limit distribution for a percolating geometry (ARC lattice)
and $\mu=0.5 < \mu_c$. The
resulting distribution shifts towards low $G$ values, while power law tails are
preserved. Log-log scales
are used so power law tails are manifest.}) and non-universal
conductivity exponents are obtained.
 For the ARC lattice we find a disorder length $\xi_D$ by
scaling the conductance of the network by $L^{\alpha}$ with
$\alpha={\ln(2/5)/\ln3}$. This scale
factor corresponds to the scaling of the conductance for the ARC network
depicted in fig.\fTwoA, with
no randomness present.  Fig.\fig\fEight{Conductance curves versus system length
for the ARC lattice with
$\mu>\mu_c$. The collapse is achieved using $\xi_D= A_+(\mu-\mu_c)^{-z'}$,
where $z'=3.2\pm 0.1$
$A_+\sim 500$. On the vertical axis $\alpha =\ln(2/5)/\ln3$.}\ shows evidence
of a
disorder length by collapsing curves for different
values of $\mu$. The horizontal axis is scaled by $\xi_D=
A_+(\mu-\mu_c)^{-z'}$. By collapsing the
curves (see fig.\fEight) we find $\mu_c=0.75\pm 0.05$ and $z'=3.2\pm 0.1$. The
value of $\mu_c$ is
close to that reported by MGM on the basis of numerical estimates. The exponent
$z'$ on the other
hand is a {\it new exponent}. Presumably this new length scale determines a
substructure of the ARC
lattice supporting most of the current. The prefactor $A_+\sim 500$, is large
in comparison to the
ones obtained above the threshold and is indicative of long crossover effects
well recognized in the
literature\ref\rTrema{A.-M. S. Tremblay and J. Machta, Phys. Rev. B {\bf 40},
5131 (1989) .}. Regarding the
scaling of the average conductance at the disorder length ($G_{\xi_D}$), we
find a very weak but
clear $\mu$ dependence as $G_{\xi_D}\sim B\log\mu$.

Below $\mu_c$, $\xi_D$ remains infinite and MDP tails are preserved through the
scales as reported
before\rMacIII\ref\rAM{R. F. Angulo and E. Medina, Physica A {\bf 191}, 410
(1992).}. We call this
behavior ``{\it Levy-like}" for its similarities to the case of addition of
random variables, when the
tail strength is such that the first and second moments diverge. The additive
limit theorem is
obviously equivalent to the resistor network in one dimension (in terms of
resistors).

\bigskip
{\noindent{\secfont II. Disorder Length}}

In the last section the existence of a disorder length scale $\xi_D$ for
homogenization was
shown numerically. In the present section we discuss the origin of this length
scale, for networks
above the threshold, starting
from the theory of Ambegaokar, Halperin and Langer\rAHL. This theory was
originally applied to transport processes that involve quantum mechanical
tunneling
or thermal activation over a barrier, where the barrier distribution is itself
broad.
The relevant distribution for these systems is of the form
\eqn\eExpD{P(g)={1\over \lambda g}~~~{\rm with}~~ 1<g<e^{\lambda},}
where the range for $g$ is restricted so that the distribution is normalizable.
This PDF is so
broad as $\lambda \rightarrow \infty$,  that conduction is restricted to a
small subset
of the whole supporting structure. AHL conjectured that the relevant subnetwork
would be dominated by the bottle-neck resistor which first establishes a
conducting path
by the following procedure: remove all conductors recalling their location of
origin,
then put them back in order of decreasing conductance. The bond laid at the
point of
percolation will determine the macroscopic conductance. This process is
formally
expressed by the {\it percolation condition}
\eqn\ePerC{\int_{g_c}^{\infty} P(g) dg = p_c,}
where $g_c$ is the governing conductance. This argument is applicable always
that $\lambda$ is
large, so that $P(g)$ in eq.\eExpD\ has strong enough tails. In the following
we shall establish,
by plausible arguments, the dependence of the disorder length $\xi_D$ on the
tail strength parameter
$\mu$. Arguments will pertain mostly to the results above the percolation
threshold described in section I.

By definition the correlation length is
\eqn\eSHK{\xi = l_0(p-p_c)^{-\nu}=l_0\lambda^{\nu}\left (\ln (g_c/g)\right
)^{-\nu},}
where we have used eq.\eExpD\ and the percolation condition. The prefactor
$\l_0\lambda^{\nu}$ defines a length
scale which we identify with the disorder length $\xi_D$. The argument builds
on a relation that
suggests a use the critical path analysis for more general distributions\rTH,
namely
\eqn\eTycHal{P(g_c)g_c = \lambda^{-1}<< 1,}
which simply states that $\lambda$ should be large. Applying the
percolation condition for the power law distributions in
eq.\ePDFg\ one arrives at the following dependence for $g_c$,
$$g_c = p_c^{1/\mu} \qquad {\rm for}~~\mu<0.$$
Using the condition in equation \eTycHal\ yields
$$p_c g_c = |\mu|g_c^{\mu}= |\mu|p_c = \lambda^{-1} \qquad {\rm for}~~\mu<0.$$
By eq.\eSHK\ one can establish that $\lambda= (\xi_D/l_0)^{1/\nu}$, so for
$\mu<0$
$$|\mu|p_c= (\xi_D/l_0)^{-1/\nu},$$
from which
\eqn\eDisL{\xi^{(-)}_D= {l_0\over p_c^{\nu}}|\mu|^{-\nu}.}
Analogously for $\mu>0$ one gets
$$\xi^{(+)}_D={l_0\over (1-p_c)^{\nu}}|\mu|^{-\nu}.$$
{}From these expressions we can also derive the ratio of the amplitudes (not
universal) which comes to be
\eqn\eRat{A_+/A_-= \left[{ (1-p_c)\over p_c}\right]^{-\nu}.}
Taking the values for $p_c$ and $\nu$, which can be computed exactly on these
lattices, we get
$A_+/A_-=0.46$ in reasonable agreement with the value $0.6$ found numerically.
This simple argument leads to the conclusion that $z = \nu$ the percolation
correlation
length exponent, and the correspondence $\mu \leftrightarrow p$. One can also
arrive at the same conclusion by a simpler
derivation as follows: From $\xi=l_0(p-p_c)^{-\nu}$ and using eqs.(1-4), one
obtains $\xi=l_0(g^{\mu}-g_c^{\mu})^{-\nu}$.
As we are interested in the $\mu\rightarrow 0$ limit we expand $g^{\mu}$ in
powers of $\mu$ and find to first order
$$\xi = \l_0 \left |{\ln g \over \ln g_c}\right |^{-\nu} |\mu|^{-\nu}$$
from which the same length scale $\xi_D$ is identified. One can alternatively
state the relation
between the AHL distribution springing from $r=r_0\exp(\lambda \epsilon)$ and
the distributions given
by eq.\ePDFg\ by choosing $\epsilon$ from the appropriate distribution
$D(\epsilon)$\rTH.

One can readily explain  the scaling variable for the conductance in figure
\fThree, by
invoking a relation first suggested by Ambegaokar,  and Kurkijarvi\ref\rACK{V.
Ambegaokar, S.
Cochran, and J. Kurkijarvi, Phys. Rev. B {\bf 8}, 3682 (1973). See also E.
Charlaix, E. Guyon, and S.
Roux , in {\it Transport in Porous Media 2} (Riedel, Dordrecht, 1987). }\ and
later discussed in
general dimensions by ${\rm Ty\check{c}}$ and Halperin\rTH,
\eqn\eTH{\langle G_{\xi_D}\rangle=Cg_c[g_c P(g_c)]^{(d-2)\nu},}
where $g_c$ is the critical conductance as if $\lambda \rightarrow \infty$,
$P(g)$ is a broad
distribution and $\nu$ is the correlation length exponent for ordinary
percolation. From
eq.\eTycHal\ the conductance $\langle G_{\xi_D}\rangle$ depends on $\lambda$.
In $d=2$ no $\lambda$
dependence is present and AHL arguments have no corrections. As a matter of
fact on square lattices
it can be shown by duality arguments that $C=1$ in eq.\eTH\ and AHL arguments
are exact. For the
general power law distribution in eq.\ePDFg\ one gets,
\eqn\ePDFg{\langle G_{\xi_D}\rangle=\cases{ C_1(1-p_c)^{1/\mu} &for $d=2$\cr
C_2(1-p_c)^{\nu+1/\mu}
\mu^{\nu} &for $d=3$\cr}}
valid for $\mu >0$. Similar formulae (with $(1-p_c) \rightarrow p_c$) hold for
$\mu<0$. In general dimensions we expect
\eqn\eScale{\langle G_{\xi_D}\rangle=\cases{(1-p_c)^tF_1({L/ \mu^{-\nu}}) &for
$\mu >0$\cr
p_c^tF_2({L/ \mu^{-\nu}}) &for $\mu < 0$\cr}}
with $t=(d-2)\nu +1/\mu$ the non-universal conductivity exponent, and $F_1$ and
$F_2$, universal functions.
This is precisely the behavior verified by collapsing data in two and three
dimensions for
$\mu \rightarrow 0$. When $\mu$ is larger, these
predictions are expected to deteriorate as the volume within $\xi_D$ gets
smaller and the AHL
percolation arguments include further corrections.

The previous discussion has only involved networks above the percolation
threshold. We have no similar
understanding for phenomena ocurring at the threshold. Most likely, the
behavior observed in this
range will follow from a renormalization group treatment similar to that of
Lubensky and Tremblay\rLT.
This treatment correctly shows (with corrections due to Machta\rMacII) that
there is
a crossover from normal conductivity exponents to non-universal ones at $\mu_c$
for lattices at the threshold.
What transpires from the numerical results is that the percolating lattice with
strong disorder
for $\mu>\mu_c$ self averages after a length scale $\xi_D$. Above this length
scale the system
behaves as if no disorder were present i.e. the length dependence of the
conductivity is purely of
geometrical origin.  As $\mu$ approaches $\mu_c$, this length diverges
giving way to a non-self-averaging situation, which explains physically the
origin of
non-universal exponents.
\eject

{\noindent{\secfont III. Discussion}}

Usually one says a system is self averaging if one can divide it in subsystems
which are
representative samples of the whole. The size of the subsystems is determined
by the distance over
which correlations persist. When such correlations persist to all scales the
system is no longer self
averaging. In this work we have caracterized how the correlations diverge for
random resistor
networks as a function of the disorder strength as measured by the broadness of
the MPD. The limit in
which correlations diverge depends on whether the  geometrical support is
critical or not. While for
lattices above the threshold only exponentially broad distributions induce
diverging correlations,
lattices at the threshold show infinite correlation at a critical broadness
given by $\mu_c$.
The disorder length $\xi_D$ discussed in the previous sections is a measure of
such correlations. The
study of this length scale has shown that its critical behavior is analogous to
that of ordinary
geometrical percolation given that $p \leftrightarrow \mu$. The length scale
$\xi_D$ is
in fact well known qualitatively in many contexts: In the study of
porous media the {\it porosity} does not acquire its intensive
macroscopic value until one measures beyond  a given length scale. This effect
is expected to
influence the scaling of other transport related quantities\ref\rKT{A. J. Katz
and A. H. Thompson,
Phys. Rev. B {\bf 34}, 8179 (1986).}\ref\rPLDa{Pierre Le Doussal, Phys. Rev. B
{\bf 39}, 4816
(1989)}\ such as the absolute and relative permeability of crucial relevance in
oil recovery. In the
context of ore mining the length scale is known as the ``nugget effect'' and is
easily measurable
(see ref.\ref\rNugget{T. Bourbie, O. Coussy and B. Zinszner, {\it Acoustics in
Porous Media},
Editions Technip (1987)}.) The knowledge of this length is very important to
the scaling theories of
these systems, generally described in terms of mean  field type
approaches\ref\rKirk{S. Kirkpatrick,
Rev. Mod. Phys {\bf 45}, 574 (1973).}. One of the most important
drawbacks of the mean field descriptions is the correct assesment of the
fluctuation correlation
length. On the other hand if by an alternate theory one can access the correct
correlation length,
one could then incorporate such information in the original mean field scheme
to improve results. One way to
do this is to obtain the renormalized distribution beyond the crossover length
(or merely its mean
and variance) and use it as the MPD in the mean field self-consistent equation.
This procedure  has
been carried out by MonteCarlo on regular lattices\ref\rATAM{A. Aponte,
P. G. Toledo, R. F. Angulo and E. Medina, unpublished.}\ with good results.

In the following we will briefly discuss other disordered systems that exhibit
analogous features
to RRN with a broad distribution of bond strengths. The simplest disordered
system with non-trivial properties
is maybe a random walker, whose displacement is given by a addition of random
variables. The limit PDF of a
sum of random variables is well known to obey the central limit theorem always
that the first two moments
of the added variables exist. When the distribution of the individual variables
is broad enough their
first and second moment diverge the Gaussian basin is not approached. In its
place a Levy basin arises in
which the limit distributions preserve the tails of the underlying random
variables. In the language of
eq.\ePDFg\ $\mu_c=2$ is the critical value of the tails that separate the
Gaussian basin from the Levy
basin. As implied by our results there is a strong parallelism between limit
theorems for addition of random variables and
RRN for lattices at $p_c$ (see Section II) where $\mu_c=0.75$, or lattices
above $p_c$ with $\mu_c=0$.
Derrida\ref\rDerr{B. Derrida, Phys. Rep. {\bf
103}, 29 (1984).}\ has established a similar analogy to spin glasses. In that
model the high temperature phase
is characterized by a peaked distribution (annealed limit) well described by
its average, while the
low temperature phase (spin glass phase) corresponds to a broad distribution of
properties. For the
Random Energy Model\ref\rDerrREM{B. Derrida, Phys. Rev. B {\bf 24}, 2613
(1981).} (the simplest Spin-Glass)
the correspondence $\mu \leftrightarrow T/T_c$ can be
established\ref\rDerriBegRohu{B. Derrida in
{\it Ecole de Physique de la Matie\`re Condense\'e}, Beg-Rohu (1993), lecture
notes.}, where
$T_c$ is the critical temperature for the transition. Presumably the same will
be true for the more
complicated Sherrington-Kirkpatrick\ref\rSK{D. Sherrington and S. Kirkpatrick,
Phys. Rev. Lett.
 {\bf 35}, 1792 (1975).} model.

Furthermore a similar phenomenon occurs in the context of directed
polymers\ref\Cook{J. Cook and B.
Derrida, J. Stat. Phys. {\bf 57}, 89 (1989).}\  where the composition law for
random variables is sums of products.
Recently Zhang\ref\rZhang{Y. -C. Zhang, Physica A {\bf 170}, 1 (1991).}\ found
that non-universal exponents
occur in the context of {\it directed polymers} if one introduces a broad
distribution of
random variables. This type of disorder presumably arises, in the context of
evolving fluid interfaces,
due to the power law nature of the breaking process that generates a porous
medium. This model
was also studied by Roux et al\rRHSLP, and the
results are very similar to ours; namely the existence of a percolating limit
(where AHL
type arguments apply well) as the broadness of the distribution increases
towards exponentially wide, and an analog of a `Levy basin'. Roux et al are
also able to find the asymptotic distributions and scaling behavior for the
simple limit
of the {\it optimal paths}. It would be interesting to study similar ``disorder
lengths" and their
divergence in these closely related systems.

As a final point we discuss the action of the {\it composition law} for random
variables as a ``filter''
for broad tails. From the results of the previous sections the stable limit PDF
can be inferred
(by inspection) from the knowledge of the microscopic PDF and the manner in
which the composition law
preserves, reduces or enhances the tail of a probability distribution. We will
illustrate this point
by discussing the simple model of Invasion Percolation. In the context of this
model, the interplay
between composition law and PDF yields an explanation for percolating backbone
nature of the
invading fluid. According to the rules of invasion percolation a non-wetting
fluid will invade a porous medium
characterized by a distribution of pore radii $Q(r)$ (or equivalently by a
distribution of
capillary pressures $K(P)$, where  $r\sim 1/P$), invading one pore at a time
and following
the path of least capillary resistance. For a hierarchical lattice such as the
one shown
in fig.\fOneA\ the composition law to produce the equivalent capillary pressure
of a cell is
\eqn\eLast{P({\rm Equivalent})
= max[min(P_{1},P_{2}), min(P_{3},P_{4})].}
The $P_i$ stand for the capillary pressures assigned to the bonds of the
hierarchical lattice.
 Such a composition law will progressively
maximize the tail of any microscopic PDF, regardless of whether it is broad or
not. In fact the
former composition law is identical to the one of conductances in the limit
that
MPD is $P(g)\sim 1/g$\rPLDa. In view of the previous conclusion, the AHL
criterion
holds and the large scale equivalent capillary pressure of the porous medium
above the percolation threshold
is identical to that of the percolating cluster of smallest capillary pressure.
The same argument applies for
wetting fluids replacing $min \rightarrow max$ and viceversa in eq.\eLast.

\vfill
\break

\centerline{\secfont Acknowledgments}

One of the authors (EM) acknowledges helpful discussions with Amnon Aharony and
Bernard Derrida.
We thank M. Araujo and P. G. Toledo for carefully reading the manuscript. The
authors thank INTEVEP S.A. for
permission to publish this paper.

\vfill\eject\immediate\closeout\rfile
\centerline{{\bf References}}\bigskip
{\catcode`\@=11\escapechar=`  \input refs.tmp\vfill\eject}
\vfill\eject\immediate\closeout\ffile
\centerline{{\bf Figure Captions}}\bigskip
{\catcode`\@=11\escapechar=`  \input figs.tmp\vfill\eject}
\end